\documentclass[aps,twocolumn,showpacs,superscriptaddress]{revtex4}
\usepackage{graphicx}

\begin{document}

\title{Modified  superexchange model for  electron tunneling across the terminated molecular wire}
\author{Elmar G. Petrov}
\affiliation{Bogolyubov Institute for Theoretical
Physics, National Academy of Sciences of Ukraine,
Metrologichna Street 14-B, UA-03680 Kiev, Ukraine}

\begin{abstract}

The explicit expressions for a nonresonant tunneling current mediated by the bridging units of the molecular wire embedded between the metallic electrodes, are derived.
The specific regimes of the charge transmission controlled by nonresonant and resonant participation of terminal's energy levels are studied, and the role of energy position of delocalized orbitals in formation of a distant superexchange electrode-electrode coupling is clarified. The criteria for reduction of the superexchange model of charge tunneling to the flat barrier model are formulated and the parameters of the barrier model (energy gap and effective electron mass) are specified in the terms of inter-site coupling and energy distance from the Fermi level to the delocalized wire's HOMO (LUMO) level. Special attention is payed to  derivation of explicit analytic expressions for the tunneling current using different approximations. It is shown that if terminal units play barrier's role in the  charge transmission process, the best correspondence with the observed current-voltage characteristics is achieved with the simplest Gauss and the mean-value explicit forms. This is supported by comparison of the theory with experimental data concerning  the current-voltage characteristics of $N-$alkanedithiol chain.

\end{abstract}

\pacs{05.60.Gg, 73.63.Nm, 85.65.+h}

\maketitle

\section{Introduction}

The use of tunnel and atomic force microscopes to study charge transport processes in molecular compounds, as well as to create molecular structures with specified conductive properties has revealed wide possibilities to utilize these structures as basic elements of molecular electronics, optoelectronics and spintronics \cite{wold01,nitz01,galp07,han02,req16,arad13,jia13,asw09,
song08,ram14,rat13,xiwa16,bag17,zha15,cap15}
Of great importance is  the mechanism of formation of the nonresonant interelectrode current in the  molecular junction "left electrode - molecular wire - right electrode" (LWR system) where a molecular wire comprises a regular chain anchored to the electrodes through its terminal units. In the nonresonant charge transmission regime, where the MOs of the wire is not occupied by the transferred electrons/holes,  the current decays exponentially with an increase in the molecular length \cite{sel02,cui02jpc,cui02,eng04,fan06,sim13,wie13}.

Analysis of $I-V$ characteristics of molecular wires is mainly performed with the simple flat-barrier Simmons model \cite{simm63}. The model predicts an exponential decrease in the tunneling current where attenuation factor  $\beta$  is expressed via two fitting parameters, the effective mass  $m^*$,  and the height of rectangular  barrier $\Delta E$. Detail analysis of the Simmons model shows \cite{sel02,cui02jpc,eng04} that a choice of the above mentioned fitting parameters, especially $\Delta E$, depends on the precise voltage region and chain length. Thus, for molecular junctions, the rectangular barrier model  does not have the uniform parameters. The model of superexchange  tunneling through a molecular wire provides an alternative approach based on mutual overlap of wave functions of the bridging interior wire units  and the overlap between the terminal wire units and the electrodes; this gives rise to formation of a direct distant coupling between the conductive states of the  spaced electrodes. The development of the superexchange model originates from pioneer McConnel's paper
\cite{mcc61} on donor-acceptor transfer of an electron through the chain of aromatic free radicals. Later the model  was expanded
for the analysis of distant donor-acceptor electron transport
through various types of organic bridging structures and protein chains  (see examples in  refs.
\cite{kha78,pet79,lar81,ber87,new91,voi13}).
McConnell's version of superexchange model \cite{mcc61} was used to describe a distant  hole transfer through DNA molecules \cite{jor02,bix02,tre02}, to analyze the $I-V$ characteristics of alkane chains \cite{eng04,ram02}, and to study the combined hopping-tunneling electron transmission in the terminated molecular wires \cite{pzmh07}.  The model explains an exponential drop of a current with the wire length but there  is a discrepancy in the attenuation factor predicted by the barrier model. In the superexchange model, the attenuation factor is determined through both  the hopping matrix element between the  sites of electron/hole localization on the neighboring units of a regular chain and  the energy distance of the Fermi level with respect to position of the \emph{localized} molecular orbitals (MOs) belonging
the interior wire unit. This energy distance  differs strongly on  barrier height $\Delta E$, which, in the case of molecular junction, is assumed to be the gap between the Fermi level and the \emph{delocalized} HOMO level belonging to the regular range of the wire \cite{cui02jpc,eng04}.

In this paper, the theory of nonresonant electron/hole superexchange tunneling is modified to derive explicit expressions for the current through a molecular junction.
Within this modified model, the expression for  attenuation factor $\beta$ yields two  different limits corresponding either the flat-barrier or McConnel's models.
The paper is organized as follows. In Section II, the tight binding model is used to derive a modified  form for the nonresonant  superexchange tunneling current along with its explicit analytic forms. Results concerning the applicability of the modified superexchange model to description of experimental data in the terminated molecular wires are presented in Section III. Concluding remarks concentrate the attention on the  explicit forms for the tunneling current as well as on corresponding the modified superexchange model to the superexchange model of a deep tunneling and the barrier model.

\section{Model and basic equations}.

\begin{figure}
\includegraphics[width=8cm]{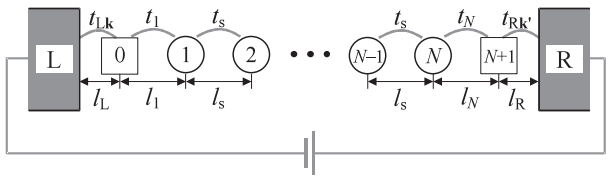}
\caption{Arrangement  of  units of  a linear molecular wire relative to the attached electrodes L and R. Explanation in the text.
}
\label{fig1}
\end{figure}

We consider  a  molecular junction  as the quantum LWR  system where a linear molecular wire is attached to the left (L) and the right (R) electrodes, Fig. \ref{fig1}.

\subsection{Hamiltonian}

The standard form for the Hamiltonian of  molecular junction reads
%
\begin{equation}
H=H_{E} + H_M +V_{E-M} \,
\label{hlmr1}
\end{equation}
where
%
\begin{equation}
H_{E}= \sum_{r=L,R}\sum_{{\bf k},\sigma}E_{r \bf k\sigma} a^{+}_{r
{\bf k}\sigma} a_{r {\bf k}\sigma} \label{rl1}
\end{equation}
is the  Hamiltonian of electrodes with  $E_{r\bf k\sigma}$ being the energy of an electron with the $\sigma$th spin projection in the $\bf k$th conduction band state of the $r$th electrode. Operators of creation  and annihilation of an electron in the single-electron band state $\bf k$  are denoted through $a^{+}_{r{\bf k}\sigma}$ and $a_{r{\bf k}\sigma}$, respectively. For a molecular Hamiltonian, we use the tight-binding model where  the transferred electron can leave  the twofold filled energy level of the \emph{highest occupied molecular orbital} (HOMO)$_n$ or occupy an empty energy level of the \emph{lowest unoccupied molecular orbital }(LUMO)$_n$ located on the wire unit $n = (0.1,...N,N+1)$.
The distance $l_{n n \pm 1}\equiv  l_s$ between the  neighboring units is associated with that of  between the sites of main electron localization within the unit. For instance, in the
(--CH$_2$--)$_N$  alkane chain, the  $l_s$ refers to a distance between the neighboring  C--C bonds.
With introduction of the creation operator $c^+_{n}$, and  the annihilation operator $c_{n}$, of an electron on the  $n$th wire unit, the molecular Hamiltonian can be represented in the following form
%
\begin{displaymath}
H_{M}= \sum_{n=0}^{N+1}\,\sum_{\sigma}\big[E_{n} c^{+}_{n\sigma} c_{n\sigma}
\end{displaymath}
\begin{equation}
+ (1-\delta_{n,N+1})\,t_{n, n+1}
 \, (c^{+}_{n\sigma} c_{nn+1\sigma} + c^{+}_{n+1\sigma} c_{n\sigma})\big]
\label{m1}
\end{equation}
In Eq. (\ref{m1}),  $E_n$ is the energy of an electron  on the $n$th unit in the presence of the bias voltage  $V=(\mu_L - \mu_R)/|e|$ where $|e|$  is the absolute value of electron charge.
For definiteness sake, the left electrode is assumed to be grounded so that chemical potential of the $r$th electrode appears as
%
\begin{equation}
\mu_{r} = E_{F} -  |e|V\delta_{r,R}\,,\;\;\;(r=L.R)\,
\label{lrchp}
\end{equation}
with $E_F$ being the energy of electrode's Fermi level.
Introducing the zero bias orbital energy  $E^{(0)}_{n}$, in a linear approximation over the $V$,  one obtains
%
\begin{equation}
E_{n} = E^{(0)}_{n} - \eta_n |e|V\,.
\label{en1}
\end{equation}
Here, the Stark shifts are characterized by the factors
$\eta_{L(R)} = l_{L(R)}/l$ at $n=0(N+1)$  and $\eta_n = [l_L+l_1 + (n-1)l_s]/l$ at $n=1,2,...N$, with  $l=l_L+l_1 + (N-1)l_s + l_{N} + l_R$ being the total  interelectrode distance. Electron couplings between the MOs of the neighboring wire units are characterized by the hopping matrix elements $t_{n, n+1}$. For  the  interior (regular) range of  a molecular wire we set  $t_{n, n+1}\equiv -t_s$ whereas $t_{0, 1}\equiv -t_1$ and  $t_{N, N+1}\equiv -t_{N}$  are used for the terminal units, Fig. \ref{fig1}.

Interaction of the chain  with the electrodes is provided by its terminal units so that the third term in the Hamiltonian (\ref{hlmr1}) appears as
%
\begin{displaymath}
V_{E-M}\,=\,\sum_{n}\,\sum_{r{\bf k}\sigma}\,(\delta_{r,L}\delta_{n,0} +\delta_{r,R}\delta_{n,N+1})\,
\end{displaymath}
\begin{equation}
\times (t_{nr {\bf k}}\,c^+_{n\sigma}a_{r {\bf k}\sigma}
+ t^*_{nr {\bf k}}\, a^{+}_{r {\bf k}\sigma}c_{n\sigma})\,.
\label{hld-m}
\end{equation}
Here, $t_{nr {\bf k}}$ is the hopping matrix element that characterizes the coupling between the $r{\bf k}$th  conductive band level and the single-electron level belonging  the MO of the $n$th terminal site.
The Hamiltonians in expressions (\ref{hlmr1}) - (\ref{hld-m}),  refer to a typical LWR systems where the terminal unit energies $E_{0}$ and $E_{N+1}$  differ from the rest unit energies $E_n$. [For example, the N--alkanethiolate  molecular wire that is  anchored to the gold contacts via the terminal sulfur atoms.]
In such LWR systems,  the  mixing between the MOs that belong to each terminal unit $n= (0, N+1)$ and the MOs related to the interior wire units $n = 1,2,...N$ is so small  that localization of terminal MOs is conserved during the electron/hole transmission across the wire.

\subsection{The wire transmission function}

Because about 98$\%$   of  tunneling current is concentrated in the elastic component \cite{troi07}, we consider  formation of  the nonresonant tunneling current associated exclusively with this component. According to the Landauer-B\"utteker approach \cite{dat95,tian98,muj02}, the elastic tunneling current is given by expression
%
\begin{equation}
I = \frac{|e|}{\pi\hbar}\int^{\mu_L}_{\mu_R}dE\, T(E, V)\,.
\label{is}
\end{equation}
Here, $T(E,V) = tr\big[\hat{G}(E)
\hat{\Gamma}^{(L)}(E)\hat{G}^{+}(E)
\hat{\Gamma^{(R)}}(E)\big]$
is the transmission function of a molecular junction where
$\hat{G}(E) = (E - \tilde{H}_M  )^{-1}$ is the Green operator  with $\tilde{H}_M = H_M + \hat{\Sigma}_L(E)  + \hat{\Sigma}_R(E)  $ being the modified molecular Hamiltonian. The modification  is caused by  molecule - electrodes  interaction (\ref{hld-m}) and is included  in the self - energy operators $\hat{\Sigma}_r(E)$. The latter specify also the width operators $\hat{\Gamma}_r(E)  = 2{\rm Im}\hat{\Sigma}_r(E)$.
In the lowest order of perturbation  in the molecule-electrodes interaction (\ref{hld-m}), only the diagonal elements of the operator $\hat{\Sigma}_r(E)$  are important for  specification of the  $\hat{\Gamma}_r(E)$ and $\hat{G}(E)$ (see more details in refs. \cite{plt05,pet06}).  Therefore,  one can set $\Gamma_r^{(\nu\nu')}(E) \approx \delta_{\nu\nu'}\Gamma_r^{(\nu)}(E)$ and
$G^{(\nu\nu')}_r(E) \approx \delta_{\nu,\nu'}\{E- {\mathcal E}_{\nu} + (i/2)[\Gamma_L^{(\nu)}(E) + \Gamma_R^{(\nu)}(E)]\}^{-1}$ where ${\mathcal E}_{\nu} $ is the eigenvalue of  the molecular Hamiltonian $H_M$ and $\Gamma_r^{(\nu)}(E) = \Gamma_r(E)[\delta_{r,L}|U_{\nu 0}|^2 + \delta_{r,R}|U_{\nu N+1}|^2]$  is the broadening of the $\nu$th wire MO.  The latter is expressed through  two kinds of quantities. The first one, $U_{\nu n}$,  refers to the elements of matrix $\hat{U}$ that with use of the transform $c_{\nu\sigma} = \sum_{n=0}^{N+1} U_{\nu n}c_{n\sigma}$  reduces the  Hamiltonian (\ref{m1}) to the diagonal form $H_M = \sum_{\nu=0}^{N+1} {\mathcal E}_{\nu} c^{+}_{\nu\sigma} c_{\nu\sigma}$. The second one is the width parameter $\Gamma_r(E) = 2\pi \sum_{\bf k} |t_{nr {\bf k}}|^2\delta(E - E_{r\bf k})\,(\delta_{r,L}\delta_{n,0} +\delta_{r,R}\delta_{n,N+1})\,$  that characterizes  broadening of the terminal MO caused by  interaction of this MO with the attached $r$th electrode. In the case of  electrodes fabricated from noble metals,   the dependence of the  width parameters  on  the transmission energy $E$ becomes unessential \cite{nitz01}.

Thus, based on the above relations and setting $\Gamma_r^{(\nu)}(E)\approx \Gamma_r^{(\nu)} $, one can express the transmission function as
%
\begin{equation}
T(E,V) = \Gamma_{L}\Gamma_{R}\Big|\sum_{\nu =0}^{N+1}\frac{U_{\nu 0} U^*_{\nu N+1}}{E- \tilde{{\mathcal E}}_{\nu}}\Big|^2\,
\label{trfa}
\end{equation}
where $\tilde{{\mathcal E}}_{\nu} ={\mathcal E}_{\nu} -i(\Gamma_L^{(\nu)} + \Gamma_R^{(\nu)})/2$ is the proper energy of the modified molecular Hamiltonian $\tilde{H}_M$.
In the LWR  system under consideration, the mixting between each terminal MO  and  interior wire MOs is assumed to be so small that when studying the nonresonant charge tunneling, one can set
$\Gamma_L^{(\nu)}\approx \Gamma_L\delta_{\nu,0}$,
$\Gamma_R^{(\nu)}\approx \Gamma_R\delta_{\nu,N+1}$ and
$\Gamma_{L(R)}^{(\nu)}\approx 0$, ($\nu \neq 0,N+1$). Here,
$\tilde{\mathcal E}_{\nu = 0(N+1)} \approx E_{0(N+1)}- i\Gamma_{L(R)}/2$ and  $\tilde{\mathcal E}_{\nu} \approx {\mathcal E}_{\nu}$, ($\nu=1,2,...N$).  ${\mathcal E}_{\nu}$ is  the eigenvalue of  those  part of Hamiltonian (\ref{m1}) which includes  the interior wire units  $n=1,2,...N$.
Diagonalization of this part  and, thus, finding the
${\mathcal E}_{\nu}$ is performed with the transform-matrix $\hat{U}^{(reg)}$. Taking into consideration that the
relation
$\sum_{\nu = 1}^{N}U^{(reg)}_{\nu 1} U^{(reg)*}_{\nu N}/(E- {\mathcal E}_{\nu}) = t_s^{N-1}/\prod_{\nu =1}^{N}(E - {\mathcal E}_{\nu})$ is satisfied   for a linear chain independently on precise form of the elements $U^{(reg)}_{\nu n}$ \cite{plt05,ptdg95}, let us rewrite the transmission function as
%
\begin{displaymath}
 T(E,V)
 \simeq \frac{\Gamma_{L}\Gamma_{R}(t_{1}t_N/t_s)^2}{[(E- E_{0})^2 +\Gamma^2_L/4][(E- E_{N+1})^2 +\Gamma^2_{R}/4]}\,
\end{displaymath}
\begin{equation}
\times T _{reg}(E,N)
\label{ddf1}
\end{equation}
where
%
\begin{equation}
 T _{reg}(E,N) = \prod_{\nu =1}^{N}\Big(\frac{t_s}{E - {\mathcal E}_{\nu}}\Big)^2\,.
\label{ddfreg}
\end{equation}
is the transmission function of a regular chain. Below, the explicit analytic form for the $T _{reg}(E,N)$ is derived for the molecular wires where  chain MOs conserve their delocalization.

\begin{figure}
\includegraphics[width=8cm]{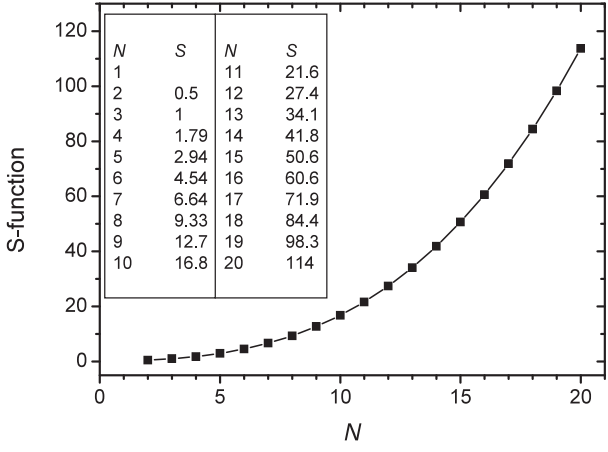}
\caption{Behavior of the auxiliary function $S(N)$, Eq. (\ref{inedel2}) v.s. the number of chain units $N$.
}
\label{fig2}
\end{figure}

Delocalization of the transferred electron/hole over the interior wire range occurs at weak bias voltages $V$ or/and at strong site-site couplings $t_s$. At $V = 0$, the transform-matrix $\hat{U}^{(reg)}$ is determined by its elements $U^{(reg)}_{\nu n} = [2/(N+1)]^{1/2}\sin{[\pi n\nu/(N+1)]}$.
Perturbation caused by the applied bias voltage is concentrated in the values  $\lambda_{\nu\,\nu'} = -|e|V\,\sum_{n=1}^N \eta_n\,U^{(reg)}_{\nu n}U^{(reg)*}_{\nu' n}$. Here, the  diagonal elements $\lambda_{\nu\,\nu} = - |e|V\eta_{c.g.}$,  determine Stark's shift of the "center of gravity" for  electron density distributed over each delocalized MO. Position of the "center of gravity"  is defined by the voltage division factor
%
\begin{equation}
\eta_{c.g.} = [l_L + l_1 + l_s(N-1)/2]/l\,.
\label{cg}
\end{equation}
The off-diagonal elements,
%
\begin{displaymath}
\lambda_{\nu\nu'} = - |e|V\,\Big(\frac{l_{s}}{l}\Big)\,
\Big[\frac{1-(-1)^{\nu + \nu'}}{2(N+1)}\Big]
\end{displaymath}
\begin{equation}
\times\Big[\frac{1}{1-\cos{\big(\frac{(\nu -\nu')\pi}{N+1}}\big)} - \frac{1}{1-\cos{\big(\frac{(\nu +\nu')\pi}{N+1}}\big)}\Big]\,,
\label{tmm}
\end{equation}
characterize the transitions between the delocalized MOs. Therefore, value
%
\begin{equation}
{\mathcal E}_{\nu} = E_{c.g.} - 2|t_s|\cos{\big(\frac{\pi\nu}{N+1}\big)}\,,\;\;\; (\nu = 1,2,...N),
\label{endel}
\end{equation}
can be referred as to  energy of the delocalized $\nu$th MO, i.e. HOMO$_{reg}$ , (HOMO -1)$_{reg}$,.. or LUMO$_{reg}$, (LUMO+1)$_{reg}$,..  if  only  the inequality
$\zeta_{\nu\nu'} \equiv |\lambda_{\nu\nu'}/({\mathcal E}_{\nu} - {\mathcal E}_{\nu'})| \ll 1$ is held for $\nu'\neq\nu$.
Dependence  of the ${\mathcal E}_{\nu}$  on the bias voltage is comprised in Stark's shift of the "center of gravity":
%
\begin{equation}
 E_{c.g.}  = E_s^{(0)} - |e|V\eta_{c.g.}\,.
\label{shdel}
\end{equation}
It should be particularly emphasized that the shift is identical for each  energy level related to the delocalized MO. The estimations show that among  values $\zeta_{\nu\nu'}$  the greatest are $\zeta_{\frac{N}{2}\, \frac{N}{2}\pm 1}$  (even $N$),  $\zeta_{\frac{N+1}{2}\, \frac{N+1}{2}\pm 1}$ (odd $N$), and $\zeta_{12} = \zeta_{N-1\,N} \equiv \zeta$ (even and odd $N$). Thus, the form (\ref{endel}) is valid if
%
\begin{equation}
\zeta = |\Delta_s/2t_s|\, S(N) \ll 1\,.
\label{inedel}
\end{equation}
Inequality (\ref{inedel}) shows that the energy drop between the identical neighboring units,
%
\begin{equation}
\Delta_s = |e|V(l_s/l)\,,
\label{ds}
\end{equation}
and the function
%
\begin{displaymath}
 S(N) =
 \Big(\frac{1}{N+1}\Big)\Big[\frac{1}{1-\cos{\big(\frac{\pi}{N+1}}\big)} - \frac{1}{1-\cos{\big(\frac{3\pi}{N+1}}\big)}\Big]
 \end{displaymath}
 \begin{equation}
 \times\Big[\frac{1}{\cos{\big(\frac{\pi}{N+1}}\big) - \cos{\big(\frac{2\pi}{N+1}}\big)}\Big] \,
\label{inedel2}
\end{equation}
are those quantities that  control  an impact of the chain length on applicability of the model of the delocalized  MOs even though $V\neq 0$.  Note that the function $S(N)$  depends solely on the number of chain units, Fig. \ref{fig2} and, thus, is identical for any regular chain.

If inequality (\ref{inedel}) is true then substitution  the ${\mathcal E}_{\nu}$,  Eq. (\ref{endel}) for the Eq. (\ref{ddfreg}) yields (see also \cite{ptdg95})
%
\begin{equation}
  T_{reg}(\epsilon,N) = \frac{\sinh^2{[\beta (\epsilon )/2]}}{\sinh^2{[(N+1)\beta(\epsilon)/2]}}\,.
\label{ddfreg1}
\end{equation}
Here,
%
\begin{equation}
  \beta(\epsilon) = 2\ln{\big[(\epsilon/2|t_s|)+\sqrt{(\epsilon/2|t_s|)^2 - 1}\big]}
 \label{df}
\end{equation}
is the attenuation factor (per chain unit) that characterizes a decrease of the chain transmission function  depending on the number of chain units $N$.  In  Eq.  (\ref{df}),
one has to substitute $\epsilon = E - E_{c.g.} > 0$ or $\epsilon = E_{c.g.} - E > 0$ if electron tunneling is mediated by the delocalized HOMOs or LUMOs, respectively. Below, for the purpose of definiteness, let us set $\mu_L\geq\mu_R$,  whereby  tunneling energy $E = E_{tun}$ lies mainly within the voltage window $\mu_{L}\geq E \geq \mu_{R}$. This means that for the  HOMO pathway the transmission occurs in the range
\begin{figure}
\includegraphics[width=8cm]{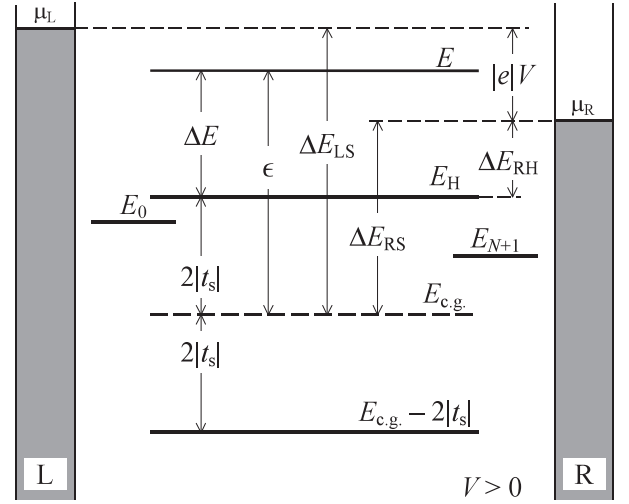}
\caption{Position of  the transmission energy $E$ in the bias voltage window at the HOMO pathway. Quantity $\Delta E(\ll 2|t_s|)$ corresponds to  the apparent barrier at superexchange  pre-resonant tunneling.
}
\label{fig3}
\end{figure}
%
%
\begin{equation}
\Delta E_{Ls} \geq \epsilon \geq \Delta E_{Rs}\,.
\label{vw}
\end{equation}
Here,
$\Delta E_{rs} = \mu_r - E_{c.g.}$ is  the energy gap  between the Fermi level of the $r$th electrode and the energy, Eq. (\ref{shdel})  associated with the "center of gravity" for the HOMOs  (cf. Fig.\ref{fig3}) .  The explicit form for the gaps is
%
\begin{equation}
\Delta E_{rs} = \Delta E^{(0)}_{s} + |e|V\,
[\eta_{c.g.}\delta_{r,L} - (1 - \eta_{c.g.})\delta_{r,R}]\,
\label{xes}
\end{equation}
where  quantity $\Delta E_s^{(0)} = E_F - E_{s}^{(0)} > 0$ is the main energy gap in an unbiased LWR.  Since the energies of  delocalized HOMO (LUMO) are arranged below (above)  the Fermi levels, then,   along with the condition  (\ref{vw}), the inequality
%
\begin{equation}
\epsilon > 2|t_s|
\label{ets}
\end{equation}
has to be satisfied at the nonresonant tunneling.

If  $\exp{[-(N+1)\beta(\epsilon)]}\ll 1$, then the dependence of the $T_{reg}(\epsilon,N)$ on the number of chain units appears in the form
%
\begin{displaymath}
  T_{reg}(\epsilon,N) \simeq   \,(t_s/\epsilon)^2
\end{displaymath}
\begin{equation}
\times\big[\delta_{N,1}+ (1- \delta_{N,1})\big(1 - {\rm e}^{-2\beta(\epsilon)}\big)^2{\rm e}^{-\beta(\epsilon)(N - 1)}\big]\,.
\label{ddfreg2}
\end{equation}
At  small site-site coupling $t_s$, when
%
\begin{equation}
  (2t_s/\epsilon)^2 \ll 1\,,
\label{mcine}
\end{equation}
Eq. (\ref{ddfreg2}) reduces to
%
\begin{equation}
  T_{reg}(\epsilon,N)\simeq\ \,(t_s/\epsilon)^{2N}\,.
\label{ddfmc}
\end{equation}
This expression reflects a superexchange version of the \emph{deep} electron tunneling originally proposed by Mc Connell for description of  distant donor-acceptor electron transfer mediated by a molecular chain \cite{mcc61}.
The corrresponding attenuation factor reads
%
\begin{equation}
\beta(\epsilon) \simeq 2\ln{(\epsilon/|t_s|)}\,.
\label{bmc}
\end{equation}
This result has also been received in ref. \cite{oni98}.
Another limiting case  happens when the energy gap  between the transmission energies $E$ and the energy $E_H = E_{c.g} + 2|t_s|$, (cf. Fig. \ref{fig3}), is small in comparison with the  $2|t_s|$ , i.e. at
%
\begin{equation}
\Delta E = \epsilon -2|t_s |\ll 2|t_s|\,.
\label{inrbar}
\end{equation}
In this case, following the approach that has been earlier derived for description of the donor-acceptor  electron transfer through the protein chains \cite{pet79}, one can introduce the  effective electron mass
%
\begin{equation}
 m^*= \hbar^2/2|t_s|l_s^2\,.
\label{em}
\end{equation}
Here, $l_s$ is the distance between the sites of electron localization on the neighboring units of a regular chain,  Fig. \ref{fig1}. [Compare  a standard definition of the effective electron mass, $1/m^{*} = (1/\hbar^2)(\partial^2 E/\partial k^2)$,  based on  expansion of  the band energy of an infinite chain, $E=E(k)$,  around the zero wave vector $k= 0$.] Using  definition (\ref{em}) one obtains
%
\begin{equation}
T_{reg}(\epsilon,N)\simeq (1/4)\,{\rm e}^{- (2/\hbar)\sqrt{2m^*\Delta E}\,d}\,,
 \label{ddfem}
\end{equation}
where $\Delta E$ is given by  Eq. (\ref{inrbar}) (see also Fig. \ref{fig3}).  The quantity  $d = (N-1)l_s$ is the distance between the sites of electron localization on the edge chain units $n = 1$ and $n = N$,  Fig. \ref{fig1}.
Formal introduction of an effective mass with  Eq.  (\ref{em}) is independent of the chain length  and, thus, is quite suitable for comparison of the superexchange model with a barrier model.

Due to inequality (\ref{inrbar}),  expression
(\ref{ddfem})  reflects the \emph{pre-resonant} regime of tunneling transmission. Such a regime mimics a tunneling through a rectangular barrier of the $\Delta E$ height, but it is well to bear in mind that  similar interpretation is true  at very specific condition determined by Eq. (\ref{inrbar}).

\subsection{Explicit forms for the current}

Our aim is to obtain the distinct expressions that  could be  acceptable to describe the $I-V$ characteristics of molecular wires containing regular bridging chains.
To this end, let us introduce the current unit $i_0\equiv (|e|/\pi\hbar)\times 1$eV$\approx 77,3\mu$A and rewrite expression (\ref{is}) in the form
%
\begin{equation}
I = i_0\,
\,\int_{-|e|V/2}^{+|e|V/2}\,d\xi\,T(\xi,V)\,. \label{curgel1}
\end{equation}
Here, the wire transmission function
%
\begin{equation}
 T(\xi,V) =  T_L(\Delta\epsilon_0 -\xi)T_{reg}(\Delta\epsilon_s -\xi,N) T_R(\Delta\epsilon_{N+1}- \xi)\,
\label{wtr}
\end{equation}
appears as the product of partial components
%
\begin{displaymath}
T_L(\Delta\epsilon_0 -\xi) = \frac{\Gamma_L}{t_s}\frac{t_1^2}{(\Delta\epsilon_0 - \xi)^2 +\Gamma_L^2/4}\,,
\end{displaymath}
\begin{equation}
T_R(\Delta\epsilon_{N+1}- \xi) = \frac{\Gamma_R}{t_s}\frac{t_N^2}{(\Delta\epsilon_{N+1}- \xi)^2 +\Gamma_R^2/4}\,,
\label{ttr}
\end{equation}
and
%
\begin{equation}
T_{reg}(\Delta\epsilon_s -\xi,N) = \frac{\sinh^2{[\beta(\Delta\epsilon_s -\xi)/2]}}{\sinh^2{[(N+1)\beta(\Delta\epsilon_s - \xi)/2]}}\,.
\label{ctr}
\end{equation}
(In the last expression, the $\beta$ - factor is given by  Eq. (\ref{df}) at $\epsilon = \Delta\epsilon_s - \xi$). If  superexchange electron tunneling is mediated by the HOMOs, then a voltage dependence  is concentrated in the quantities
%
\begin{displaymath}
\Delta\epsilon_0 =\Delta E_0^{(0)} - (|e|V/2)(1-2\eta_L)\,,
\end{displaymath}
\begin{equation}
\Delta\epsilon_{N+1} =\Delta E_{N+1}^{(0)} + (|e|V/2)(1-2\eta_R)\,,
\label{x1n}
\end{equation}
and
%
\begin{equation}
\Delta\epsilon_s = \Delta E_s^{(0)} -(|e|V/2)(1-2\eta_{c.g.})\,.
\label{xes}
\end{equation}
The transmission functions
exhibit a specific voltage behavior dependently on position of the gaps  (\ref{x1n}) and (\ref{xes})  relative to  energy window
%
\begin{equation}
-|e|V/2\geq \xi\geq |e|V/2\,.
\label{wxi}
\end{equation}

At nonresonant tunneling, the gap
$\Delta\epsilon_s $ remains always outside this window. Moreover, at perfectly symmetric molecular junction, i.e. at $\eta_{c.g.} = 1/2$,  it becomes independent of the $V$ so that $\Delta\epsilon_s = \Delta E_s^{(0)}$. As to the rest
gaps, they can enter in the energy window either at $V > 0$ (gap $\Delta\epsilon_{0}$) or at $V < 0$ (gap $\Delta\epsilon_{N+1}$).
In a LWR  system where  $\Delta\epsilon_{0} > |e|V/2$,   the $T(\xi,V)$ increases smoothly with  $\xi$ . It is not the case if  $\Delta\epsilon_{0} <  |e|V/2$ where  $T(\xi,V)$ manifests the presence of a peak  at $\xi = \Delta\epsilon_0$ associated with the partial component $T_L(\Delta\epsilon_0 -\xi)$. This circumstance allows one to express the current, Eq. (\ref{curgel1}) with three possible explicit forms.

\subsubsection{Zero Gauss approximation}

The simplest form is associated with  zero Gauss approximation, when $ T(\xi, V)$ is taken at the middle of the integration limits, i.e. at $\xi = 0$. This yields
%
\begin{equation}
I\approx I_G  = i_0\,|e|V T_L(\Delta\epsilon_0)
T_{reg}(\Delta\epsilon_s,N) T_R(\Delta\epsilon_{N+1})\,,
\label{igauss}
\end{equation}
Introducing the current mediated by a single bridging unit,
%
\begin{equation}
I_{G}(1) = i_0\,|e|V\,
\frac{\Gamma_L\Gamma_Rt_1^2t_N^2}{\Delta\epsilon_s^2[(\Delta \epsilon_{0})^2 +
\Gamma_L^2/4][(\Delta\epsilon_{N+1})^2 +\Gamma_R^2/4]}\,,
\label{iunit}
\end{equation}
one obtains
%
\begin{equation}
I_{G} = I_{G}(1)\,\Phi (\beta_s,N)\,.                                                                                                                                                                                                                                                                                                                                                     \label{igauss1}
\end{equation}
Distant behavior of  current  $I_G$ is comprised in
the chain attenuation function
%
\begin{equation}
\Phi (\beta_s,N)  =
\frac{\sinh^2{\beta_s}}{\sinh^2{[(N+1)(\beta_s/2)]}} \,,                                                                                                                                                                                                                                                                                                                                                     \label{lf}
\end{equation}
that is equil to unit at $N=1$. Corresponding  attenuation factor $\beta_s$  is given by  Eq. (\ref{df}) at $ \epsilon = \Delta\epsilon_s$. For a chain where  $\exp{[- (N+1)\beta_s]}\ll 1$, the drop appears as
%
\begin{equation}
\Phi (\beta_s,N)  \approx \big(1 - {\rm e}^{-2\beta_s}\big)^2{\rm e}^{-\beta_s(N - 1)}\,.                                                                                                                                                                                                                                                                                                                                                     \label{lfr}
\end{equation}

\subsubsection{Mean-value approximation}

To obtain the second explicit form for the current,  we employ the mean-value (M.V.) approximation, whereby the transmission functions  (\ref{ttr}) and   (\ref{ctr})  are substituted for  averaged  values  $\overline{T}_{L(R)}$ and  $\overline{T}_{reg}$, respectively. Thus,
%
\begin{equation}
I_{M.V.} = i_0\, |e|V \,\overline{T}_L\overline{T}_{reg}(N)
\overline{T}_R\,
\label{imv}
\end{equation}
where
%
\begin{equation}
\overline{T}_L = \frac{2\,t_1^2}{t_s|e|V}\Big[{\rm tan}^{-1}{\Big(\frac{2\Delta E_{L0}}{\Gamma_L}\Big)} - {\rm tan}^{-1}{\Big(\frac{2\Delta E_{R0}}{\Gamma_L}\Big)} \Big]\,
\label{tlav}
\end{equation}
and
%
\begin{equation}
\overline{T}_R = \frac{2\,t_N^2}{t_s|e|V}\Big[{\rm tan}^{-1}{\Big(\frac{2\Delta E_{L\,N+1}}{\Gamma_R}\Big)} - {\rm tan}^{-1}{\Big(\frac{2\Delta E_{R\,N+1}}{\Gamma_R}\Big)} \Big]\,.
\label{trav}
\end{equation}
Averaging is performed within the window (\ref{wxi}).
Voltage dependence of each function $\overline{T}_{r}$,  ($r=L,R$),  is deduced in terminal gaps $\Delta E_{rn} =\mu_r - E_n$ that  read
%
\begin{displaymath}
\Delta E_{r0} = \Delta E_0^{(0)} + |e|V[\eta_L\,\delta_{r,L} - (1-\eta_L)\,\delta_{r,R}]\,,
\end{displaymath}
\begin{equation}
\Delta E_{rN+1} = \Delta E_{N+1}^{(0)} - |e|V[\eta_R\,\delta_{r,R} + (1-\eta_R)\,\delta_{r,L}]\,,
\label{tergap}
\end{equation}
Note the abrupt behavior of   $\overline{T}_{r}$  in the vicinity of those $V$ where  the gaps  $\Delta E_{rn}$ change their sign. If $|\Delta E_{rn}| \gg \Gamma_r$, then
for estimation of the  $\overline{T}_r$ one can use the approximation
%
\begin{equation}
{\rm tan}^{-1}{\Big(\frac{2\Delta E_{rn}}{\Gamma_r}\Big)\approx \frac{\pi}{2}\Big[\frac{\Delta E_{rn}}{|\Delta E_{rn}|} - \frac{\Gamma_r}{\pi\Delta E_{rn}}} \Big]\,.
\label{tg}
\end{equation}

The expressions for  $\overline{T}_{reg}(N)$
with $N=1$ and $N=2$ bridging units are respectively
%
\begin{equation}
\overline{T}_{reg}(N = 1) = \frac{t_s^2}{\Delta\epsilon_s^2 - (|e|V/2)^2}\,
\label{att1}
\end{equation}
and
%
\begin{displaymath}
\overline{T}_{reg}(N = 2)
\end{displaymath}
\begin{displaymath}
= \frac{t_s^2}{4}\Big\{ \Big[
\frac{1}{(\Delta\epsilon_s - t_s)^2  - (|e|V/2)^2} +
\frac{1}{(\Delta\epsilon_s + t_s)^2  - (|e|V/2)^2}
\Big]
\end{displaymath}
\begin{equation}
+ \frac{1}{|e|Vt_s}\ln{\Big[ \frac{\Delta\epsilon_s^2 - (t_s + |e|V/2)^2}{\Delta\epsilon_s^2 - (t_s - |e|V/2)^2}
\Big]}\Big\} \,.
\label{att2}
\end{equation}
When the number of  chain units  exceeds 2, then
with a high degree of precision,
%
\begin{displaymath}
\overline{T}_{reg}(N\geq 3) \simeq \Big(\frac{t_s}{|e|V}\Big)\frac{1}{2N-1}
\end{displaymath}
\begin{equation}
\times\Big[F(\beta_R){\rm e}^{- \beta_R[N-(1/2)]}
- F(\beta_L){\rm e}^{- \beta_L[N-(1/2)]} \Big]\,
\label{trregav}
\end{equation}
where
%
\begin{displaymath}
F(\beta) = 1 - (2N-1)\Big[\frac{3}{2N+1}{\rm e}^{-\beta} + \frac{3}{2N+3}{\rm e}^{-2\beta}
\end{displaymath}
\begin{equation}
+ \frac{1}{2N+5}{\rm e}^{-3\beta}\Big]\,.
\label{f}
\end{equation}
Voltage dependence of quantity  $\overline{T}_{reg}(N\geq 3)$ is concentrated in the corresponding attenuation factors $\beta_{L}$  and $\beta_{R}$  determined by the Eq.  (\ref{df}) with $\epsilon = \Delta E_{L(R)s}$, Eq. (\ref{xes}).

At a near-bias voltage, above expressions reduce to
%
\begin{displaymath}
\overline{T}_{reg}(N = 1)  \approx
\Big(\frac{t_s}{\Delta E_s^{(0)}}\Big)^2\,,
\end{displaymath}
\begin{displaymath}
\overline{T}_{reg}(N = 2)  \approx
\frac{t_s}{\sqrt{(\Delta E_s^{(0)})^2  - 4t_s^2}}
\end{displaymath}
\begin{equation}
\times\frac{(2t_s)^3}{\Big(\Delta E_s^{(0)} +\sqrt{(\Delta E_s^{(0)})^2  - 4t_s^2}\Big)^3}
\label{att12s}
\end{equation}
and
%
\begin{equation}
\overline{T}_{reg}(N\geq 3) \approx \frac{t_s(1 - {\rm e}^{- \beta_0})^3}{\sqrt{(\Delta E_s^{(0)})^2  - 4t_s^2}}{\rm e}^{- \beta_0 [N-(1/2)]}\,.
\label{nb3}
\end{equation}
Quantity $\beta_0$ is the zero-bias attenuation factor for a regular chain. Its general and McConnel's forms are  given by respective  Eqs. (\ref{df}) and (\ref{bmc}) at $\epsilon = \Delta E_s^{(0)}$.

If, independently of a  bias voltage magnitude,  the terminal energies $E_0$ and $E_{N+1}$  remain below the Fermi level of electrodes, then the gaps  (\ref{tergap}) are positive  independently of the  polarity. This simplifies expressions for $\overline{T}_r$ yielding
%
\begin{displaymath}
I\approx I_{M.V.} = i_0\, |e|V \,\frac{\Gamma_L\Gamma_R}{t_s^2}
\frac{t_1^2t_N^2}{\Delta E_{L0}\Delta E_{LN+1}\Delta E_{R0}\Delta E_{RN+1}}
\end{displaymath}
\begin{equation}
\times\overline{T}_{reg}(N)\,.
\label{imv2}
\end{equation}

\subsubsection{Terminal unit approximation}

If at certain bias voltages, the gaps  (\ref{x1n}) fall within energy window (\ref{wxi}), then the terminal transmission functions (\ref{ttr})  exhibit a maximum at  $\xi = \Delta \epsilon_{0(N+1)}$. This corresponds to the transmission process at which the tunneling energy $E$ enters in resonance with the respective  energies $E_0$  and $E_{N+1}$  of terminal units. The third possible explicit version for the tunneling current supposes that the transmission functions of a regular chain as well as one of the terminal units are extracted from the integral (\ref{curgel1}) at $\xi = \Delta \epsilon_{0}$  or $\xi =  \Delta \epsilon_{N+1}$ depending on the voltage polarity. A similar simplification could be referred to the terminal unit approximation (T.U.) that appears in the form
%
\begin{displaymath}
I\approx I_{T.U.} = i_0\, |e|V\,\big[\overline{T}_L  T_{reg}(\Delta E_{0s}, N)T_R(\Delta E_{0N+1})\,\Theta(V)
\end{displaymath}
\begin{equation}
+ T_L(-\Delta E_{0N+1})T_{reg}(\Delta E_{N+1s}, N)\overline{T}_R\,\Theta(-V)\big]\,.
\label{ires}
\end{equation}
Here, energy  gaps
%
\begin{equation}
\Delta E_{0s} = E_{0}^{(0)} - E_{s}^{(0)} + |e|V(\eta_{c.g.} - \eta_L)\,
\label{g0s}
\end{equation}
and
%
\begin{equation}
\Delta E_{N+1s} = E_{N+1}^{(0)} - E_{s}^{(0)} - |e|V(1 - \eta_{c.g.} - \eta_R)\,
\label{gN+1s}
\end{equation}
coincide with the
energy distances between the terminal HOMO$_0$  and  HOMO$_{N+1}$  levels and the position of "center of gravity" of  interior  wire units, Fig. \ref{fig3}. As to an energy distance
$\Delta E_{0N+1}= E_{0} - E_{N+1}$ between the terminal  levels, it reads
%
\begin{equation}
\Delta E_{0N+1}= E_{0}^{(0)} - E_{N+1}^{(0)}
+ |e|V(1- \eta_L - \eta_R)\,.
\label{0N+1}
\end{equation}
[If  tunneling  is mediated  by the LUMOs, one can use  Eqs.  (\ref{ires}) - (\ref{gN+1s}) with substitution $\Delta E_{ns}$ for $\Delta E_{sn} = E_s - E_n$].
Note now that in line with Eqs. (\ref{tlav}), (\ref{trav}) and (\ref{tg})  one can set  $\overline{T}_{L(R)}
\approx (\pi t_{1(N)}^2/t_s|e|V$. This yields
%
\begin{displaymath}
I_{T.U.} = i_0\, \pi\,\Big\{\frac{\Gamma_Rt_1^2t_N^2}{\Delta E_{0s}^2[\Delta E_{0N+1}^2 + (\Gamma_R/2)^2]}\,
\Phi (\beta_{0s},N)\Theta(V)
\end{displaymath}
\begin{equation}
 -  \frac{\Gamma_Lt_1^2t_N^2}{\Delta E_{N+1s}^2[\Delta E_{0N+1}^2 + (\Gamma_L/2)^2]}\,\Phi (\beta_{N+1s},N)\Theta(-V)\Big\}\,.
\label{iresd}
\end{equation}
The chain attenuation  function $\Phi (\beta_{0(N+1)s},N)$ is given by  Eq. (\ref{lf}). The expression for the corresponding attenuation factor $\beta_{0(N+1)s}$ follows from  Eq. (\ref{df}) at $ \epsilon = \Delta E_{0(N+1)s}$.

\section{Results and discussion}

The modified  model of  superexchange tunneling under consideration works in much more soft condition (\ref{ets}) as opposed to the condition (\ref{mcine})  valid for  the deep tunneling. There are two kinds of  basic physical parameters, that specify the model:  the transmission gaps and the couplings.  Among the gaps,  the main are $\Delta E_{Ls}$  and $\Delta E_{Rs}$ ( see definition (\ref{xes}) and Fig. \ref{fig3}).  Nonresonant superexchange tunneling occurs at the condition (\ref{ets}).
Since the transmission energy $\epsilon$ enters in the window (\ref{vw}), then at positive polarity
($\mu_L > \mu_R $) the noted condition  reads $\Delta E_{Rs}  >  2|t_s| $.
When  $V$ exceeds a critical voltage
%
\begin{equation}
V_{RH} = \frac{\Delta E_{s}^{(0)} -2|t_s|}{|e|(1 - \eta_{c.g.})}\,,
\label{rh}
\end{equation}
the gap  $\Delta E_{RH} = \Delta E_{Rs} - 2|t_s|$ becomes negative and, thus,  charge  transmission  occurs at  the resonant  regime. Physically,   $V_{RH}$ is the value at which chemical potential of  right electrode $\mu_R$ and the energy $E_H  = E_{c.g.} + 2|t_s|$ (cf. Fig.\ref{fig3}).  Along with critical voltage (\ref{rh}), there exists the second critical voltage,
%
\begin{equation}
V_{R0} = \frac{\Delta E_{0}^{(0)}}{|e|(1 - \eta_{L})}\,.
\label{cvor}
\end{equation}
Expression (\ref{cvor}) results from the condition that  the gap $\Delta E_{R0}$, Eq. (\ref{tergap}) vanishes at $V = V_{R0}$.
Realization of the precise nonresonant tunneling regime is controlled by relations between above critical voltages.  If $V_{R0} > V_{RH}$, then  transmission energy $E$ is not able to enter in resonance with energy $E_0$  related to the $0$th  terminal unit and, thus, this unit plays an inactive (bridging) role in the tunneling. This case is presented in Fig.\ref{fig3}.
It is not the case if $V_{R0} < V_{RH}$. Now,  at $V = V_{R0}$, a specific  resonant transmission  becomes possible  via the localized  HOMO of the $0$th unit and, thus, the role of this unit becomes active  in the nonresonant tunneling through a regular range of the wire. (At the negative polarity, a similar  conclusion refers to  terminal unit $n = N+1$).

The theory establishes  a correspondence between the pair of  basic parameters of the modified  superexchange  model (zero bias gap $\Delta E_{0}^{(0)}$ and inter-site coupling $t_s$) from one side and  the pair of observable values (zero bias attenuation factor $\beta_0$ and critical voltage $V_{RH}$) from another side. Using  definition (\ref{df}) at $\epsilon  = \Delta E_{0}^{(0)}$ and expression (\ref{rh}) one arrives to  relations
%
\begin{equation}
\Delta E_{s}^{(0)} = |e|V_{RH}\frac{(1-\eta_{c.g.})
\cosh{(\beta_0/2)}}{2\sinh^2{(\beta_0/4)}^2}
\label{es}
\end{equation}
and
%
\begin{equation}
|t_s| = |e|V_{RH}\frac{1-\eta_{c.g.}}{4\sinh^2{(\beta_0/4)}}\,,
\label{ts}
\end{equation}
which clarify essentially the analysis of charge transmission processes.

Below, to demonstrate the mechanism of formation of the nonresonant tunneling current  in molecular junctions,  we consider the simplest case of perfectly symmetric LWR  system. In such a system,
%
\begin{displaymath}
\Delta E_{N+1}^{(0)} = \Delta E_{0}^{(0)} \equiv \Delta E_{*}\,,
\end{displaymath}
\begin{displaymath}
t_{N+1} = t_1\equiv t_{*}\,,
\end{displaymath}
\begin{displaymath}
\Gamma_L = \Gamma_R \equiv \Gamma_{*}\,,
\end{displaymath}
\begin{equation}
\eta_L = \eta_R \equiv \eta_{*}\,,
\label{pslwr}
\end{equation}
and $\eta_{c.g.} = 1/2$. [The symbol $*$ is used for identical terminal units.]

\subsection{Bridging role of terminal units}

As an example, let us consider a tunneling across the
$N$ - alkanedithiol chain anchored to  gold contacts via sulfur atoms.
To simplify the analysis, we omit non principal details attributed to the differences between the actual  bond lengths between  backbone  atoms and introduce the average bond length  ${\overline a}$. This means that in accord with the Fig. {\ref{fig1}, one has to set  $l_L \approx  l_1 \approx   l_s \approx   l_N \approx  l_R \equiv {\overline a}$.  Thus,
%
\begin{displaymath}
l \approx (N+1)\overline{a}\,,
\end{displaymath}
\begin{equation}
\eta_{*} = 1/(N+3)\,.
\label{vtrm}
\end{equation}
Due to the property $I(-V)  =  - I(V)$ , it is quite sufficient to  consider the $I-V$ characteristics at the positive  polarity only.
Actual geometric position of a sulfur atom relative to  surface gold atoms is unknown a priori. Therefore, electrode-molecule couplings may noticeably differ in magnitude.
Note also that a mutual  position of terminal and interior wire orbital levels is varied depending on the calculation methods \cite{bag17,sim13}.  Besides, the  energy position of  localized and delocalized orbitals with respect to  electrode's Fermi level  is not exactly known. Therefore, we pay a particular attention to a semi-phenomenological estimation of the fitting parameters with the use of relations (\ref{es}) and (\ref{ts}). For instance, the following  correspondence,
%
\begin{equation}
 \Delta E_{s}^{(0)} = 2|t_s|\cosh{(\beta_0/2)}\,,
\label{br}
\end{equation}
exists between the zero bias gap  and intersite coupling.
In the case of  deep tunneling, a similar   correspondence follows from  Eq.  (\ref{bmc}) and appears in the form
%
\begin{equation}
 \Delta E_{s}^{(0)} = |t_s|\exp{(\beta_0/2)}\,.
\label{brmcl}
\end{equation}
For a typical value $\beta_0 =1$ per CH$_2$ unit,   relations (\ref{br})  and (\ref{brmcl})  reduce to  $\Delta E_{s}^{(0)}\approx 2.27\,t_s$  and  $\Delta E_{s}^{(0)} \approx 1.65\,t_s $, respectively. [In alkane chains, the coupling $t_s$ is positive, so that $|t_s| = t_s$]. In both models,
$\Delta E_{s}^{(0)}/ t_s\sim 1 $.  This ratio is in contradiction with condition (\ref{mcine}) of applicability of  McConnel's  model. Therefore, the model of deep tunneling meets difficulties in its application to analyze $I-V$ characteristics in  $N-$alkane's wires.

An approximate estimation of hopping integral $t_s$ can be performed using comparison of  the HOMO energies of alkane chains \cite{elk17}  with zero bias  energies $E_{H}(N) = {\mathcal E}_{\nu = N}$. The latter  are given by the Eq. (\ref{endel}) at $V =0$ and reads
%
\begin{equation}
 E_{H}(N)  = E_{s}^{(0)}  + 2|t_s|\cos{\big(\frac{\pi}{N+1}\big)}\,.
\label{endelh}
\end{equation}
As a result, approximation of   $E_{H}(N)$ by  Eq. (\ref{endelh}) is more and less adequate only for the chains with $N\geq 6$. In this case, the fitting parameters can be taken as $E_s^{(0)}\approx -12.84$ eV and $t_s\approx$ 2.97 eV. For  short chains, the correlations modify strongly both $E_s^{(0)}$  and $t_s$.
Thus, the uncertainty exists in specification of coupling $t_s$ and, owing to  relation  (\ref{br}), in finding the basic zero bias gap $\Delta E_{s}^{(0)}$. For instance, the same magnitude for the attenuation factor, $\beta_0 =1$ per C-C bond,  is  obtained at $\Delta E_{s}^{(0)}$ = 3.86, 4.99, 7.04 eV if  $t_s$ = 1.71, 2.23, 3.12 eV, respectively. The uncertainty is removed if one knows critical voltage $V_{RH}$ wherein  resonance tunneling is switched on  through  chain's delocalized orbitals of a long chain.
Therefore, both basic parameters of superexchange model, $\Delta E_{s}^{(0)}$ and $t_s$, can be simultaneously estimated with the use of expressions (\ref{es}) and (\ref{ts}).
The experimental results show that in  $N$ - alkanedithiols, the ohmic  $I-V$ characteristics are held at $|V|\approx$  0.1- 0.5 V \cite{cui02jpc,eng04,sim13,fan06} and no conductance peaks are observed outside of   1.5 V \cite{eng04}. Thus, one can suppose  that  nonresonant tunneling occurs in voltage region $ V < $1.5 V so that one can set $V_{RH} = 1.5$ V. Substituting this magnitude and $\eta_{c.g.} = 1/2$ in  Eqs. (\ref{es}) and (\ref{ts}),  one can see that value $\beta_0  \approx 1$ is obtained at $\Delta E_{s}^{(0)} \approx 6.3$ eV and  $t_s\approx$ 2.78 eV.  Value 2.78 eV  does not contradict  data presented in ref. \cite{elk17}. Moreover, a direct calculation of tight binding parameters in graphene shows that for neighboring carbon atoms $t_s\approx 2.74$ eV \cite{kun11}. [For  comparison, the magnitudes of  Au-Au and Au-S site-site couplings are
$t_{\rm Au-Au}\approx 2$eV and $t_{\rm Au-S}\approx 2.65$ eV, respectively  \cite{tian98}.]
It follows from the recent results of quantum-chemical
calculations  of orbital energies in oligoethelene glycol chains \cite{bag17} that in the (CH$_2$CH$_2$O) unit, energy  of sulfur lone pair orbitals are positioned about 2.9 eV above energy of the  C-C bonding orbital so that we can set
$E_{*}^{(0)} - E_{s}^{(0)} \approx 2.9$\,eV. Now it becomes possible to specify  zero bias energy gaps between the Fermi level and the energy levels of both delocalized HOMOs as well as localized lone pair orbitals.  They are $\Delta E_{s}^{(0)} \approx $6.3 eV and $\Delta E_{*}^{(0)}\approx$ 3.4 eV, respectively. [Obviously, for the bonding single or triple Au-S orbitals, the $\Delta E_{*}^{(0)}$ exceeds 3.4 eV]. At $V\neq 0$, the zero bias gaps are transformed into those  given by  Eqs. (\ref{xes}) and (\ref{tergap}).

The dependence of tunneling current on the number of  C-C bonds in a $N$ - alkanedithiol wire is shown in Fig. \ref{fig4}. It is seen that  the best correspondence between basic integral form for the current, Eq. (\ref{curgel1}) and its analytic versions
is achieved in the framework of the mean-value approximation. %
\begin{figure}
\includegraphics[width=8cm]{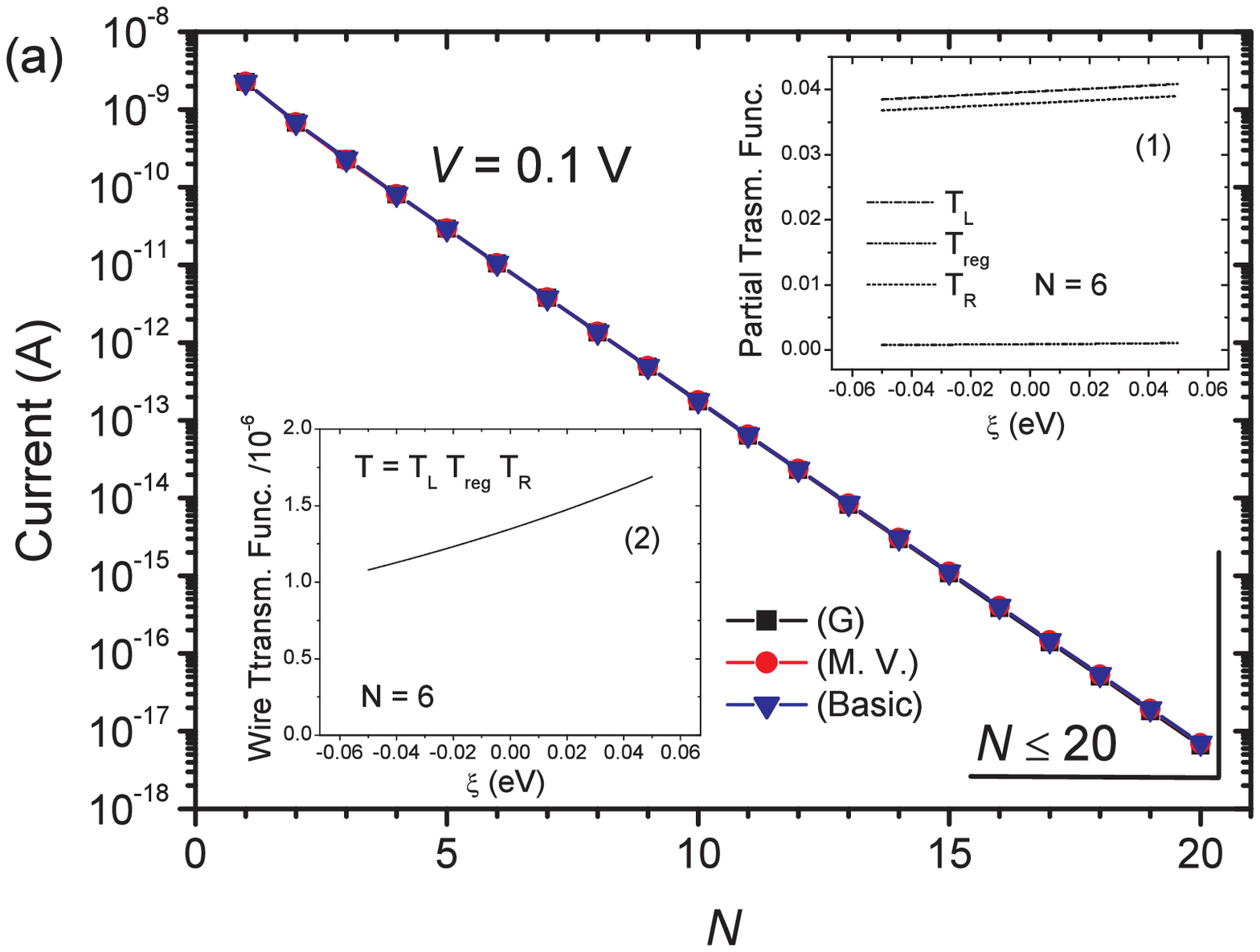}
\includegraphics[width=8cm]{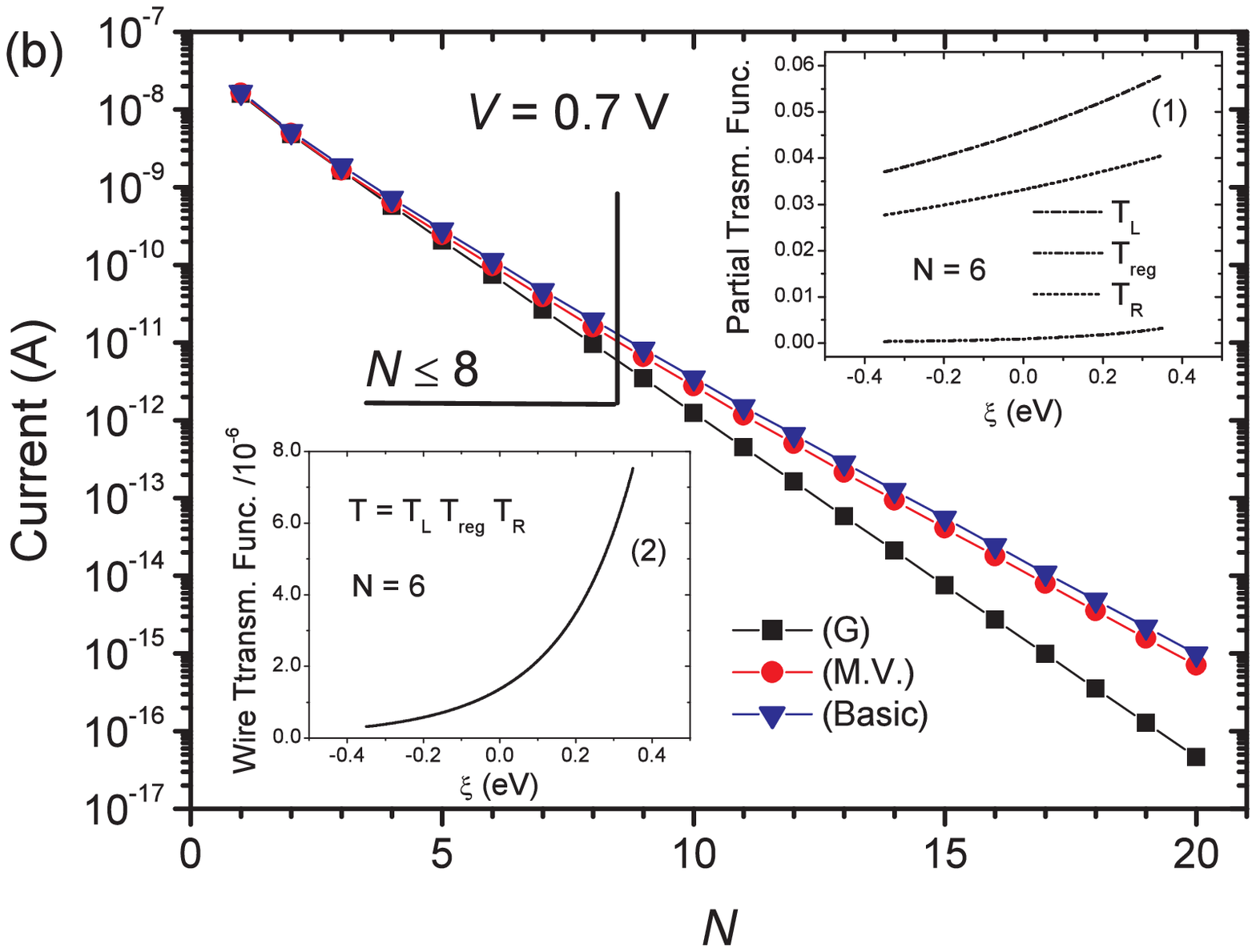}
\includegraphics[width=8cm]{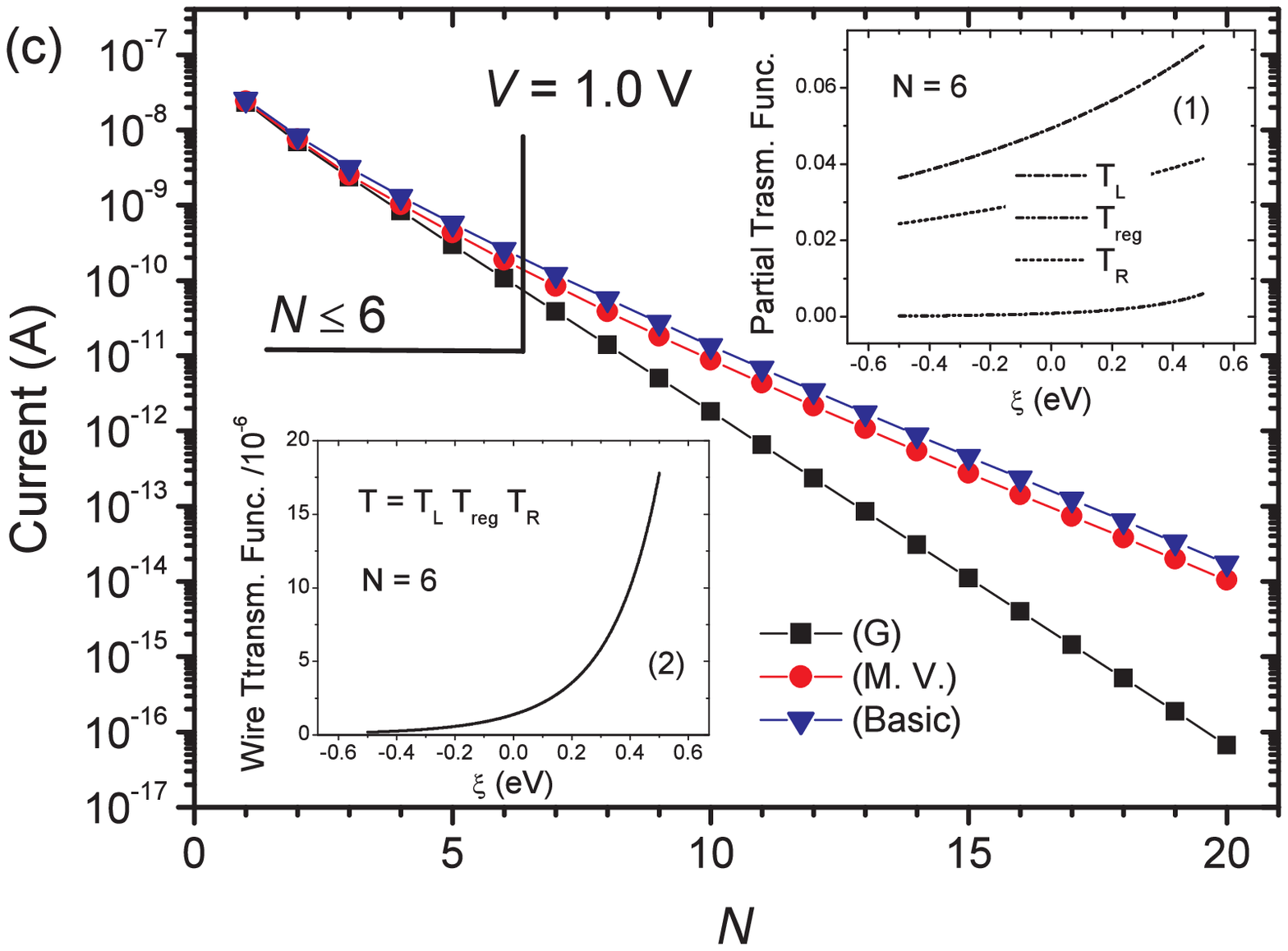}
\caption{ Attenuation of the nonresonant  tunneling current with an increase of the number of C-C bonds of $N$ - alkanedithiol molecular wire. Due to a monotonic behavior of total and partial transmission functions (cf. the insertions), the mean-value distinct form for the current, Eq. (\ref{imv})  and (\ref{imvs}) is in a good correspondence with the numerical data generated by the more exact basic integral form, Eq. (\ref{curgel1}). For bias voltages $V$= 0.1, 0.5 and 1 V, the applicability of the model  is limited by the unit numbers $N < 20, 9$ and $N < 6$, respectively. Calculation parameters  are $\Delta E_* = 3.4$ eV, $t_* = 2.50$ eV, $t_s = 2.78$ eV, $\Gamma_* = 0.2$ eV.
}
\label{fig4}
\end{figure}
This approximation brings to   Eq. (\ref{imv2}) that, for a perfectly symmetric molecular junction, reduces to the form
%
\begin{displaymath}
I_{M.V.} = i_0\, |e|V \,
\frac{(\Gamma_*t_{*}^{2}/t_s)^2}{[\Delta E_{*}^2 - (|e|V
\eta_*)^2][\Delta E_{*}^2 - (|e|V)^2(1-\eta_*)^2]}
\end{displaymath}
\begin{equation}
\times\overline{T}_{reg}(N)\,
\label{imvs}
\end{equation}
with  $\overline{T}_{reg}(N)$ specified  by  Eqs. (\ref{att1}) - (\ref{trregav}). It is important to notice that at small voltages (up to 0.2 V), the simplest Gauss approximation,
%
\begin{equation}
I_{G} = i_0\,|e|V\, \Big[
\frac{\Gamma_*t_*^2/\Delta E_s^{(0)}}
{\Delta E_{*}^2  - (|e|V/2)^2(1-2\eta_*)^2}\Big]^2\Phi(\beta_0, N)\,,
\label{iunita}
\end{equation}
provides a similar magnitude of the current as the mean-value approximation.
In Eq. (\ref{iunita}),  due to the fact that $\Delta\epsilon_s = \Delta E_s^{(0)}$ (cf. definition (\ref{xes})),  one obtains $\beta_s = \beta_0$. Thus, function $\Phi(\beta_0, N)$ is identical  to  $\Phi(\beta_s, N)$, Eq. (\ref{lf}).
As a result, the dependence of attenuation factor $\beta_0$ on the bias voltage  vanishes. Moreover, as  inequality (\ref{inrbar}) is satisfied at $\epsilon = \Delta E_s^{(0)}$, one can introduce  the effective mass, Eq. (\ref{em}). This allows one  to represent the attenuation factor (per C-C unit) in the form
%
\begin{equation}
\beta_0 = (2/\hbar)\sqrt{2m^*\Delta E}\,\overline{a}
\label{attg}
\end{equation}
where $\Delta E = \Delta E_s^{(0)} - 2 t_s \approx$ 0.74 eV.  For $\overline{a} = 1.3$ \AA, the  estimation of effective mass yields $m^*\approx 0. 8 m_e$  where $ m_e$  is the elementary electron mass. Analogously, one can introduce attenuation factors $\beta_L$ and $\beta_R$ that characterize  a decrease of $\overline{T}_{reg}(N)$, Eq. (\ref{trregav}). They read
%
\begin{equation}
\beta_{r} = (2/\hbar)\sqrt{2m^*\Delta E_{r}}\,\overline{a}
\label{attlr}
\end{equation}
where $\Delta E_{r} = \Delta E_{rs} - 2t_s$, Fig. \ref{fig3} . Bearing in mind  Eq. (\ref{xes}), this yields
%
\begin{equation}
\Delta E_{r} =  \Delta E + (|e|V/2) (\delta_{r,L} - \delta_{r,R})\,.
\label{attlr0}
\end{equation}

Recollect now that the model where superexchange tunneling is mediated by the delocalized MOs works until  perturbation caused by a bias voltage, does not destroy the delocalization. Condition (\ref{inedel}) shows that at  $t_s =2.78$ eV and
$V = 0.1$ V the delocalization is well maintained for $N\leq 20$   whereas at $V = 1.5$ V it is broken at $N > 5$.  Fig. \ref{fig5} shows  good correspondence between the theory and the experiment for those alkane chains   where, at precise  bias voltages,   inequality  (\ref{inedel})  is satisfied.
\begin{figure}
\includegraphics[width=8cm]{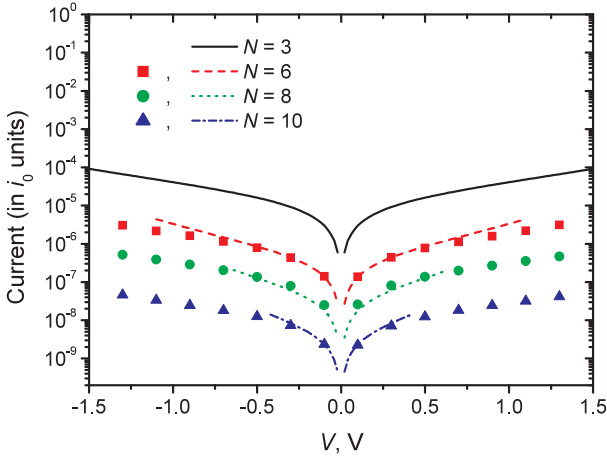}
\caption{$I-V$ characteristics of a LWR - system where the
$N - $ alkanedithiol structure fulfills the function of a molecular wire. The data points represent the  data adopted from the experimental $I = I(V)$ plots \cite{eng04}.  Each theoretical  curve covers only  limiting number of spots. This is  in correspondence with  condition (\ref{inedel}) of applicability of the modified  superexchange model of nonresonant tunneling mediated by the delocalized chain MOs (HOMOs in a given case). The parameters are the same as in Fig. \protect\ref{fig4}.
}
\label{fig5}
\end{figure}
Moreover, the theory  is able to explain why the rectangular barrier model is works at low biases but meets difficulties in its application to long chains.

\subsection{Active role of terminal units}

Let us assume that in the LWR - system,
a  regular range of the wire binds to electrodes through specific terminal units whose localized HOMOs are not far from the Fermi levels of the electrodes.
It is supposed that at certain voltages,  terminal unit's energy $E_0$ (at $V > 0$) or $E_{N+1}$  (at $V < 0$) can enter in a local resonance with the respective electrode's Fermi level. This means that at each polarity, there exists two transmission bias regions.
At the positive polarity,  these regions  are  $0 \leq V < V_{R0}$ and $V_{R0} \leq V < V_{RH}$ where critical voltages are given by the expressions  (\ref{rh}) and (\ref{cvor}). Fig. \ref{fig6} illustrates the behavior of nonresonant tunneling current as  a  function of the  number of chain units in both noted regions. The case of a perfectly symmetric %
\begin{figure}
\includegraphics[width=8cm]{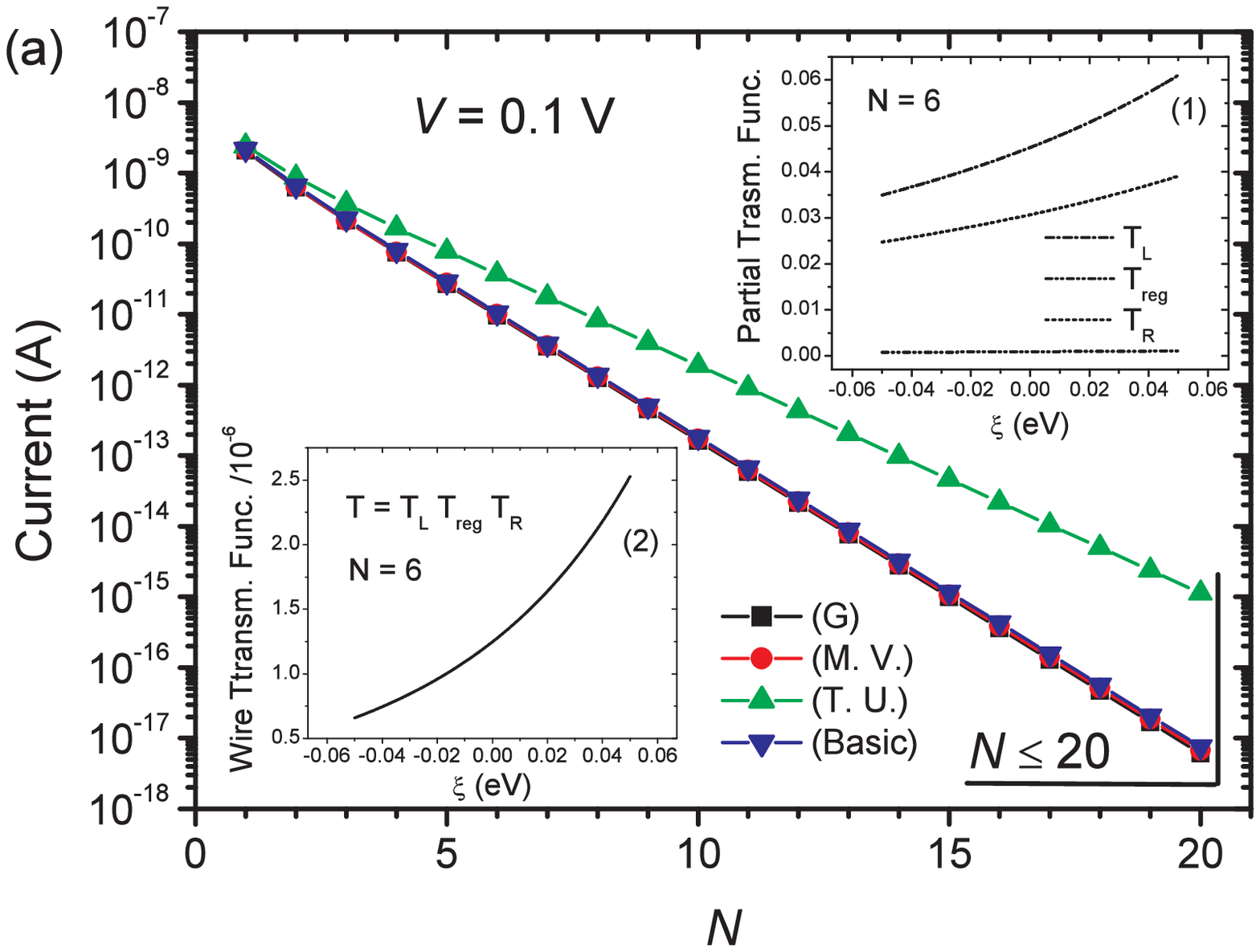}
\includegraphics[width=8cm]{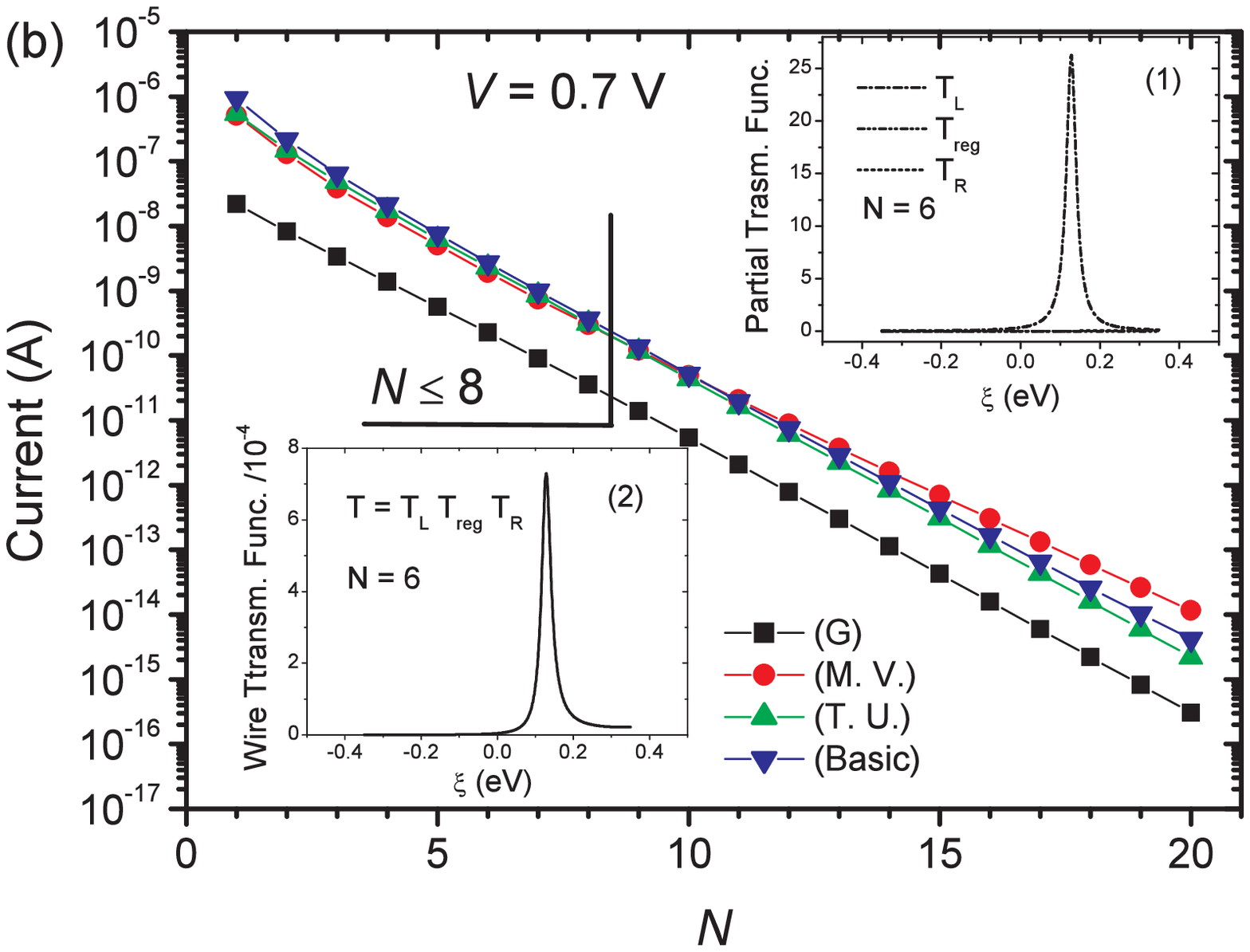}
\includegraphics[width=8cm]{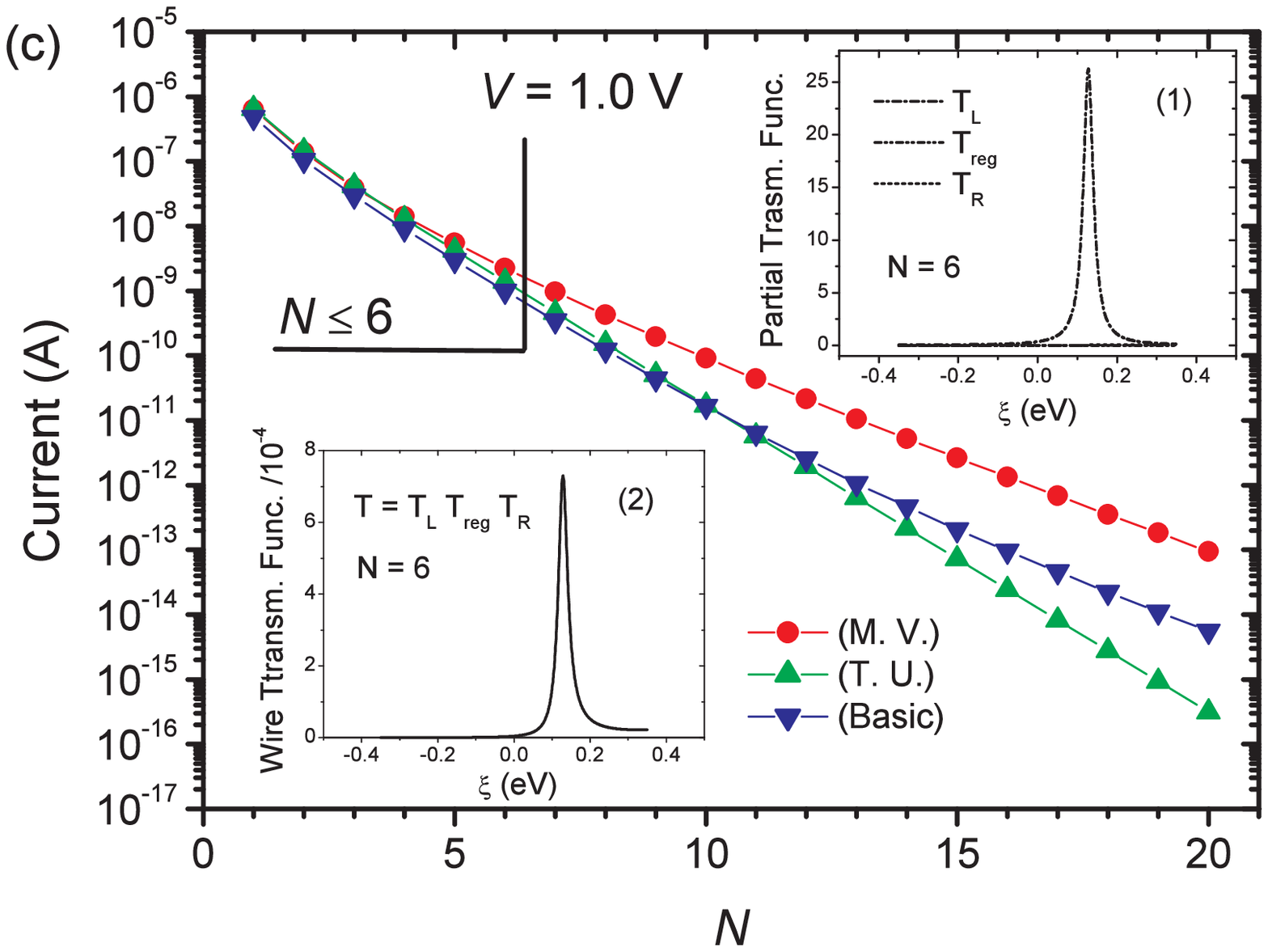}
\caption{Attenuation of the nonresonant  tunneling current with an increase of the number of C-C bonds of  $N$ - alkanedithiol wire attached  to the electrodes through  active terminal units. The mean- value approximation  form is in the best correspondence with the basic integral one. The zero Gauss  and terminal-unit explicit forms work well in voltage regions $V< V_{R0}$ and $V> V_{R0}$, respectively. The superexchange parameters of the chain are the same as in Fig. \protect\ref{fig4}. The remaining parameters are $\Delta E_* = 0.4$ eV,  $t_*$ = 0.74 eV, $\Gamma_* = 0.03$ eV. Theory works at $N\leq 20$ (at $V = 0.1$ V, (a)), $N\leq 8$ (at $V = 0.7$ V, (b)) and $N\leq 6$ (at $V = 1.0$ V, (c)).
}
\label{fig6}
\end{figure}
molecular junction is considered and the chain superexchange parameters, $\Delta E_s^{(0)}$ and $t_s$, are chosen identical to those  for the alkane chain. This allows one to use the same limitation for the applicability of the theory of superexchange tunneling mediated by the delocalized chain HOMOs.  Transmission regime in the region $ V < V_{R0}$ (Fig. \ref{fig6}a) is identical to the regime that controls a  nonresonant  tunneling with participation of the bridging terminal units (Fig. \ref{fig4}a). This is due to the fact  that at $ V =0.1$V, terminal energy $E_0$ is below the $\mu_R$ and, thus, the peak of terminal transmission function $T_L(\xi)$, Eq. (\ref{ttr}) lies outside the integration region of the basic expression for the current, Eq. (\ref{curgel1}). Therefore,  owing to a monotonic behavior of the total wire transmission function $T(\xi,V)$ in  energy window (\ref{wxi}) (see the insertions to Fig. \ref{fig6}a), both the zero Gauss and mean-value approximations give, in fact, the identical results with basic integral expression for a current, Eq. ( \ref{curgel1}) while the edge-unit approximation shows significant deviation. Situation changes essentially in region $V > V_{R0}$ where the peak of the $T_L(\xi)$ enters in energy window (\ref{wxi}) (cf. the insertions to Figs. \ref{fig6}b,c). Here, the best coincidence with the basic integral expression belongs to the explicit mean-value and terminal-unit forms. For a symmetric LWR - system, the  terminal-unit form for the current reads
%
\begin{displaymath}
T_{T.U.} = i_0\Gamma_*
\frac{(t_*^4/t_s^2)}{\Delta E_{*s}(\Delta E_{0N+1}^2 + \Gamma_*^2)}
\end{displaymath}
\begin{equation}
\times [\Phi(\beta_{0s},N)\Theta(V) -  \Phi(\beta_{N+1s},N)\Theta(-V)]\,.
\label{iress}
\end{equation}
As to Gauss's approximation, it leads to great disagreements with others (in Fig. \ref{fig6}c the Gauss approximation is omitted).

\section{Conclusions}

The main result of this paper is to obtain explicit  formulas for analyzing tunnel volt-ampere characteristics of a linear molecular wire. It is assumed that the mixing of localized orbitals of terminal units of wire with delocalized orbitals of the inner range wire (regular chain) is weak. Under such conditions the origin of the interelectrode superexchange coupling is associated with the overlap of localized electronic wave functions of terminal wire units as with band electronic wave functions of adjacent electrodes, and with delocalized wave functions of a regular chain. The superexchange coupling supposes that wire's orbital states are not populated by the transferred electron/hole and therefore participate in  formation  of the interelectrode superexchange coupling in a virtual way. The mechanism of  superexchange tunneling  through localized and delocalized orbitals remains stable only if the condition (\ref{inedel}), in which the delocalization of the orbitals of a regular chain is not destroyed by the bias voltage, is satisfied. Received explicit expressions for the nonresonance tunneling current show that under an ohmic regime of charge transmission,  it is convenient to analyze the current dependence on the number of chain units using the simplest expression for the current  obtained in the zero Gauss approximation, Eq. (\ref{igauss1}). Applicable to a perfectly symmetric molecular wire the expression for the current reduces to Eq. (\ref{iunita}), from which it follows that the
factor of exponential attenuation of the current is independent of the applied voltage until  delocalization of the orbitals of the regular chain is conserved. A more accurate mean-value approximation makes it possible to analyze
volt-ampere characteristics of a molecular wire not only for
ohmic regime (where there is a coincidence with the results of the Gauss approximation), but also outside the ohmic regime.

Comparison with experimental data (Fig. \ref{fig5}) shows that for the alkane chain, a good agreement between theoretical and experimental volt-ampere characteristics is carried out at
all those chain lengths at which the electric field does not destroy delocalization of orbitals. This result reflects one of the principal differences between the modified superexchange model from the model of a "deep" superexchange tunneling. The latter is based on  the overlapping of the wave functions of localized orbitals belonging the terminal and regular wire units. Therefore, in the "deep" tunneling model, the damping factor (\ref{bmc}) corresponds to one of the limiting cases of the expression (\ref{df}) derived in the framework of  the modified superexchange model.

Another important result of the modified superexchange  model is that when the condition (\ref{inrbar}) is satisfied,  the attenuation factor (\ref{df}) is transformed  into a form used in the phenomenological barrier model (see expressions (\ref{attg}) - (\ref{attlr0})). Thus, the modified superexchange model establishes the limit of applicability of the barrier models for the analysis of current-voltage characteristics of the molecular chains, and also connects the parameters of the barrier model (the effective tunneling mass of an electron, height and width of the barrier) with the characteristics of the molecular
chain.

The role of terminal units in a distant nonresonance tunneling is determined by the position  of the terminal energy levels  with respect to the  Fermi levels of the  adjacent electrodes.
If the orbital energies of terminal units are positioned from the Fermi levels  about  several electron volts, then the terminal units perform a role of the bridging structures, creating the tunnel barriers between  the terminal units  and the adjacent electrodes.  If, however, the energy gaps between the terminal  orbital energies and Fermi levels are such that, with the experimentally achievable bias voltages the  orbital energies enter in local resonance  with the chemical potentials of the electrodes, then terminal units begin to play an active role. This role appears in the enhancement of the distant nonresonant tunneling current by the inclusion of local resonant transmission processes between the electrodes and
adjacent terminal units (compare the magnitudes of the
current in Figs. 4 and Fig. 6 at $V = 0.7$ V  and $V = 1$ V).

In general, it can be said that the use of  the modified
superexchange model has less restrictions on conditions of applicability in comparison with the model of "deep" tunneling or the model of rectangular barrier (which most often used for the analysis of tunneling currents through molecular chains). Explicit formulas of the modified superexchange models have well-defined  physical ranges of their applicability and therefore are convenient for clarifying of  the peculiarities of the formation not only of the tunnel current, but also the conductivity and the resistance in different types of molecular wires.

\section {Acknowledgments}
The present work was partially supported by The National Academy of Sciences of Ukraine (project No. 0116U002067)


%

\begin{thebibliography}{00}
%


\bibitem{wold01} D. J. Wold and C. D. Frisbie, J. Am. Chem. Soc. \textbf{123}, 5549 (2001).

\bibitem{nitz01} A. Nitzan Ann. Rev, Phys. Chem. , \textbf{52}, 681 (2001).

\bibitem{galp07}M. Galperin, M. A. Ratner, and A. Nitzan, J. Phys. Cond. Matter, \textbf{19}, 103201 (2007).

\bibitem{han02} P. H\"{a}nggi, M. Ratner, and S. Yaliraki (eds.), Special Issue, Chem. Phys. {\bf 281}, 111 (2002).

\bibitem{req16} R. Requist, P. P. Baruselli, A. Smogunov, M. Fabrizio, S. Modesti,  and E. Tosatti, . \textbf{11}, 499 (2016)

\bibitem{arad13} S. V. Aradhya and L. Venkataraman,
Nature Nanotechnol. \textbf{8} , 399 (2013).

\bibitem{jia13} C. Jia and  X. Guo,  Chem. Soc. Rev. \textbf{42}, 5642 (2013).

\bibitem{asw09} D. K. Aswal, S. P. Koiry, B. Jousselme, S. K. Gupta, S. Palacin, and J. V. Yakhmi, Physica E \textbf{41}, 325 (2009).

\bibitem{song08} H.Song, H. Lee, and T. Lee, Ultramicroscopy
 \textbf{108}, 1196  (2008).

\bibitem{ram14} K. V. Raman, Appl. Phys. Rev. \textbf{1}, 031101 (2014).

\bibitem{rat13} M. Ratner, Nature Technology, \textbf{8}, 377 (2013).

\bibitem{xiwa16} D. Xiang, X. Wang, Ch. Jia, T. Lee, and X. Guo, Chem. Rev. \textbf{116}, 4318 (2016).

\bibitem{bag17}M. Baghbanzadeh, C. M. Bowers, D. Rappoport, T. Zaba, L. Yuan, K. Kang, K.-C. Liao, M. Gonidec, P. Rothemund, P. Cyganik, A. Aspuru-Guzik, and G. M. Whitesides, J. Am. Chem. Soc. \textbf{139}, 7624 (2017).

\bibitem{zha15} J. L. Zhang, J. Q. Zhong, J. D. Lin, W. P. Hu, K. Wu, G. Q. Xu, A. T. Wee, and W. Chen, Chem Soc Rev. \textbf{44}, 2998 (2015).

\bibitem{cap15} B. Cappozzi, J. Xia, O. Adak, E. J. Dell, Z. E. Lin, J. C. Taylor, J. B. Neaton, L. Campos, and L. Venkataraman, Natur Nanotechnology \textbf{10}, 522 (2015)

\bibitem{sel02} J. Selzer, A, Salomon, and D. Cahen, J. Phys. Chem. B \textbf{106}, 10432 (2002)

\bibitem{cui02jpc} X. D. Cui, A. Primak, X. Zarate, J. Tomfohr, O. F. Sankey,  A. L. Moore, T. A. Moore, D. Gust, L. A. Nagahara, and S. M. Lindsay, J. Phys. Chem. B, \textbf{106}, 8609 (2002).

\bibitem{cui02} X. D. Cui, X. Zarate, J. Tomfohr, O. F. Sankey,  A. Primak, A. L. Moore, T. A. Moore, D. Gust, G. Harris, S. M. Lindsay, Nanotechnology, \textbf{13}, 5 (2002).

\bibitem{eng04}V. B. Engelkes, J. M. Beebe, and C. D. Frisbie, J. Am. Chem. Soc. \textbf{126}, 14287 (2004).

\bibitem{fan06} F. Chen, X. Li, J. Hihath, Z. Huang, and N. Tao, J. Am. Chem. Soc. \textbf{128}, 15874 (2006).

\bibitem{sim13} F. C. Simeone, H. J, Yoon, M. M. Thuo, J. R. Barber, B. Smith, G. M. Whitesides, J. Am. Chem. Soc. \textbf{135}, 18131 (2013).

\bibitem{wie13} E. Wierzbinski, X. Yin, K. Werling, and D. H. Waldeck, J.Phys. Chem.B \textbf{117}, 4431 (2013).

\bibitem{simm63}J. G. Simmons, J. Appl. Phys. \textbf{34}, 1793 (1963)

\bibitem{mcc61} H. M. McConnel, J. Phys. Chem. \textbf{35}, 508 (1961).

\bibitem{kha78} V. N. Kharkyanen, E. G. Petrov, and I. I. Ukrainskii, J. Theor. Biol. \textbf{73}, 29 (1978).

\bibitem{pet79} E.G. Petrov, Int. J. Quant. Chem. \textbf{16}, 133 (1979).

\bibitem{lar81} S. Larsson, J. Am. Chem. Soc. 103, 4034 (1981).

\bibitem{ber87} D. N. Beratan, J, N, Onuchiv, and J. J. Hopfield, J. Chem. Phys. \textbf{86}, 4488 (1987).

\bibitem{new91} M. D. Newton,  Chem. Rev. \textbf{91}, 767 (1991).

\bibitem{voi13} A. A. Voityuk, J. Phys. Chem. C, \textbf{117}, 2670 (2013).

\bibitem{jor02} J. Jortner, M. Bixon, A. A. Voityuk, and N. R{\"o}sch, J. Phys. Chem. A, \textbf{108}, 7599 (2002).

\bibitem{bix02} M. Bixon and J. Jortner, Chem. Phys. \textbf{281}, 393-408 (2002).

\bibitem{tre02}C. R.Treadway, M. G. Hill,  and J. K. Barton, Chem. Phys. \textbf{281}, 409 (2002).

\bibitem{ram02}M. A. Rampi and G. M. Whitesides, Chem. Phys. \textbf{281}, 373 (2002).

\bibitem{pzmh07} E. G. Petrov, Ya. R. Zelinskyy. V. May, and P. H{\"a}nggi, J. Chem. Phys. \textbf{127}, 084709 (2007).

\bibitem{troi07} A. Troisi and M. A. Ratner, Small \textbf{2}, 172 (2007)

\bibitem{dat95}S. Datta, \emph{Electronic Transport in Mesoscopic Systems} (Cambridge University Press, Cambridge, UK, 1995).

\bibitem{tian98} W. Tian, S. Datta, S. Hong, R. Reifenberger, J. I. Henderson, and  C. P. Kubiak, J. Chem. Phys. \textbf{109}, 2874 (1998).

\bibitem{muj02} V. Mujica and M.A. Ratner, Chem. Phys. \textbf{281}, 147 (2002).

\bibitem{plt05} E.G. Petrov, Low Temp. Phys. \textbf{31} (3-4), 338 (2005).

\bibitem{pet06} E.G. Petrov, Chem. Phys. \textbf{326}, 151 (2006).

\bibitem{ptdg95} E.G.~Petrov, I.S.~Tolokh, A.A.~Demidenko,and V.V.~Gorbach, Chem. Phys. {\bf 193}, 237 (1995).

\bibitem{oni98} A. Onipko, Yu. Klymenko, L. Malysheva, and S. Stafstr\"om, Solid St. Comm. \textbf{108}, 555 (1998).

\bibitem{elk17}S. Elke, C. J. Carlos, \emph{Molecular Electronics: An Introduction in Theory and Experiment} (2nd Edition), In Nanoscience and Nanotechnology (Vol. 15). World Scientific, Singapore, 2017.

 \bibitem{kun11} R. Kundu, Mod. Phys. Lett. \textbf{25}, 163 (2011).



\end{thebibliography}
\end{document}